\documentclass[12pt]{iopart}

\usepackage{iopams} 
\usepackage{xcolor}
\usepackage{graphicx}
\begin{document}

\title{Numerical study of the effect of mass of the background gas on the lateral interactions of two plasma plumes at high pressure}

\author{Sharad K. Yadav$^{1}$\footnote{sharadyadav@iisc.ac.in;sharadky@gmail.com} and R. K. Singh$^{2}$ \footnote{rajesh@ipr.res.in}}

\address{$^{1}$Department of Physics, Indian Institute of Science (IISc.), Bangalore 560012, Karnataka, India}
\address{$^{2}$Institute for Plasma Research (IPR), Gandhinagar 382428, India}
\begin{abstract}
The characteristic of the lateral interaction of two plasma plumes in argon
$Ar$ background gas at high pressures was reported in recent publication [Yadav {\it et. 
al.}, J. Phys. D: Appl. Phys. {\bf 50}, 053421 (2017)].
Further we have investigated the interaction characteristics of plumes 
in $He$, $Ne$, $Ar$ and $Xe$ background gases to see the effect of mass on the 
interaction. The present work illustrate the applicability of the present model for theoretical 
understanding of dynamics, structure, density variation, shock wave formations and their interactions of 
two propagating plasma plumes  in a wide range of ambient conditions. The formation of interaction 
region, geometrical shape and strength of the shock fronts and subsequent regular and Mach reflections in 
accordance with the nature and pressure of ambient gas are successfully captured in the simulations.  The 
observed results are supported by the reported experimental observations under identical conditions.
\end{abstract}
\maketitle
\section{Introduction}
Laser interaction with matter and subsequent evolution of 
target material as a form of plasma plume has 
a variety of applications in many areas such as pulse laser 
deposition, nano-particles/clusters formation, material 
processing, elemental analysis, lithography, atmospheric 
plasma and plasma diagnostics
\cite{allen1995,chrisey1994,choudhury2019,miziolek2006,freeman2011,
zakharov2003,huber2005}. Many parameters such as
laser wavelength and energy density, properties of material (e.g. 
thermal conductivity, heat capacity, density) and also the 
reflectivity and absorption of material collectively govern the plasma 
formation. Hence the basic mechanism of the formation and the evolution of 
laser produced plasma plume is a complex process and its theoretical 
understanding continues to be a challenging task. \par
Apart from the extensive research on single laser plasma plume, the interaction between the plasma 
plumes, also known as colliding plasmas \cite{hough2010} has been subject of great interest because of its applications 
in  laser ion source, inertial confinement fusion (ICF), and laboratory simulation and modelling of 
astrophysical plasma phenomena \cite{bulanov2002,rancu1995,wan1997,gregory2008,kuramitsu2011,elton1994}.  When the plasma plumes interact under certain conditions, an 
interaction region or plasma jet likes structure is formed \cite{dardis2010}. The dynamics and plasma parameters of this 
additional jet like structure is depend on the geometry of the interaction and plasma parameters of 
interacting plumes (seed plumes). Several experiments have been done to manipulate the induced plasma 
jet like structure by using the different interaction geometry, e.g. head-on collisions, angular and 
lateral interactions where the seed plasma plumes are generated with wide range of laser 
intensities \cite{alshboul2014,ake2006,harilal2011,eagleton1997,luna2007,kumar2014}. The increasing 
interest in colliding plasmas induced interaction reason is largely 
due to its better control of the plasma parameters and geometrical shape in accordance with its 
applications. \par
Plasma jet produced by energetic colliding plasmas immersed as a important tool for 
laboratory scaled model of various astrophysical phenomenon \cite{gregory2008,kuramitsu2011,elton1994}. Camps {\it et. al.} \cite{camps2002}   
utilize the colliding 
plasma to produce an aggregate-free materials source. Y. Hirooka {\it et. al.} \cite{hirooka2011} study the 
Aerosol formation and hydrogen co-deposition by colliding ablation plasma plumes from plasma facing 
element. Further colliding plasmas is used to understand the stagnation on 
hohlraum axis and capsule implosions in ICF \cite{rancu1995,wan1997}. In addition to the widely explored the collision 
between the conventional laser produced plasmas, recently several experiments have been conducted to 
understand the interaction between the Laser-Blow-Off (LBO) of the thin film 
\cite{kumar2015,kumar2016}.  Due to the difference in ablation mechanism, the thermal history, 
composition and evolution of LBO plume is significantly different from plasma plume produced by bulk 
solid target \cite{singh2007}. Since the major constituents of LBO plume is neutral species and 
therefore the interaction between LBO plasmas can be used to generate the directed beam of neutral 
species for tokomak plasma diagnostics \cite{lie1984,pospieszczyk1989}. \par 
In spite of extensive application oriented work on the laser produced colliding plasma plumes, its 
theoretical understanding are scarce in the literature. Also the interaction between the plasma plumes 
in presence of ambient gas is more complex in comparison to the case of vacuum because of the presence 
of shock front ahead of the plasma plume \cite{kumuduni1993,george2013,zeldovich2002}. In presence of 
shock front, the iteration between the plumes is governed by shock-shock collision and its reflection \cite{kumar2015}. 
The understanding of colliding shock is important, especially in probing the astronomical object in 
laboratory scale \cite{gregory2008,kuramitsu2011}. \par
Recently, we have simulated the lateral interactions between 
two spatially separated LBO plasma plumes in the presence of argon ambient \cite{yadav2017}. In this 
approach, continuity, momentum and the energy equations of ablated material are solved numerically in 
two dimension \cite{yadav2017,patel2012}. In this numerical study we showed the formation of shock 
waves and their interactions in argon ambient which is in good agreement with experimental 
observations \cite{yadav2017}. Since evolution of the plasma plume and also the geometrical shape and 
strength of the shock wave is depends on the on the pressure and mass of the ambient gas; which 
collectively determine the dynamics and structure formation of the induced interaction region. 
Therefore we are motivated to look more closely the interactions between the LBO plumes in wide range 
of ambient environment. \par
In extension of our earlier work \cite{yadav2017}, here we report the two 
dimensional numerical simulation of interaction of LBO plume in $He$, $Ne$, $Ar$ and $Xe$ background. These 
background gases are chosen because of large difference in their atomic masses and the physical 
parameters. 
\section{Modeling of Plasma Plumes and Numerical Simulation}
\subsection{Basic fluid equations:} 
We modeled the time evolution of plasma-plume in the ambient gas using basic fluid equations that include 
mass, momentum and energy equations. \par
Mass equation:
\begin{eqnarray}
\frac{\partial \rho}{\partial t}=-\vec{\nabla} \cdot (\rho \vec{u})
\end{eqnarray}
In eq. $\left(1\right)$ $\rho$ is the mass density of whole system i.e. $\rho=\rho_{v}+\rho_{b}$; $
\rho_{v}$ is the plume density (also known as vapor density) and $\rho_{b}$ is the background gas density; 
$\vec{u}$ is the flow velocity. \par
Mass equation (only for the plasma plume):
\begin{eqnarray}
\frac{\partial \rho_{v}}{\partial t}=-\vec{\nabla} \cdot ( 
\rho_{v} \vec{u}) +\vec{\nabla} \cdot(\rho 
D_{mix}\vec{\nabla}\Omega_{v})
\end{eqnarray}
In which $D_{mix}$ is the binary diffusion coefficient expressed as 
$D_{mix}=\frac{2}{3}\left(\frac{k}{\pi} \right)^{3/2}\left(\frac{1}{2m_{v}}+\frac{1}{2m_{b}}
\right)^{\frac{1}{2}}\frac{T^{\frac{3}{2}}}{p\left(\frac{d_{v}+d_{b}}{2}\right)^{2}}$; $d_{v}$ and 
$d_{b}$ are the size of the vapor and background gas species respectively; pressure $p$ and 
temperature $T$ are related to each other by the relation 
$ p=\rho_{v}kT/m_{v}+\rho_{b}kT/m_{b}$;
$m_{v}$ is the mass of the vapor plume species and $m_{b}$ is the mass of the
background gas species. $\Omega_{v}$ 
is the vapor mass density fraction i.e. $\Omega_{v}=\rho_{v}/\rho$. \par
Momentum equation:
\begin{eqnarray}
\frac{\partial \rho \vec{u}}{\partial t}=-\vec{\nabla}\cdot 
(\rho \vec{u} \vec{u})-\vec{\nabla} p +\vec{\nabla}\cdot 
\bar{\bar{Q}}
\end{eqnarray} \par
Energy equation:
\begin{eqnarray}
\frac{\partial E}{\partial t}=-\vec{\nabla} \cdot (E\vec{u})-
p\vec{\nabla}\cdot \vec{u}-\vec{\nabla}\cdot \vec{q}+\bar{\bar 
{Q}}:\vec{\nabla}\vec{u}+\vec{\nabla}\cdot [(E_{v}-E_{b})\rho 
D_{mix}\vec{\nabla}\Omega_{v}]
\end{eqnarray}
We consider the transfer of momentum  due to the symmetric viscous stress tensor $\bar{\bar{Q}}$ defined as:
\begin{eqnarray}
\bar{\bar{Q}}=\eta_{mix}\left[\left\lbrace \vec{\nabla}\vec{v}+\left(\vec{\nabla}\vec{v}\right)^{'} \right\rbrace-\frac{2}{3}\left(\vec{\nabla}\cdot \vec{v}\right)\bar{\bar{I}}\right]
\end{eqnarray}
where $(\bar{\nabla}\bar{v})^{'}$ represents the transpose of the tensor $\vec{\nabla}\vec{v}$. The viscosity $\eta_{mix}$ of the binary mixture is determined using semi-empirical formula of Wilke provided in Ref. \cite{bird2006} as.
\begin{eqnarray}
\eta_{mix}=\sum_{i}\frac{f_{i}\eta_{i}}{\sum_{j} f_{j}\Phi_{ij}}
\end{eqnarray}
The suffixes $i$ and $j$ are to be summed over the two species of the vapor $v$ and background gas $b$ for the case of binary gas mixture. Here $f_{i}$ and $\eta_{i}$ are the number density fraction and the viscosity respectively for the $i^{th}$ species and $\Phi_{ij}$ is a dimensionless number given by the expression below.\\
\begin{eqnarray}
\Phi_{ij}=\frac{1}{\sqrt{8}}\left(1+\frac{m_{i}}{m_{j}}\right)^{-1/2}\left[1+\left(\frac{\eta_{i}}{\eta_{j}}\right)^{1/2}\left(\frac{m_{j}}{m_{i}}\right)^{1/4}\right]^{2}
\end{eqnarray}
The viscosity corresponding to the individual species is calculated from the kinetic theory of the gas,
\begin{eqnarray}
\eta_{i}=\frac{1}{\pi d_{i}^{2}}\sqrt{\frac{m_{i}kT}{\pi}}
\end{eqnarray}
where $E$ in eq.$\left(4\right)$ is the total internal energy $\it{i.e.}$ 
$E=\frac{3}{2}\left(\frac{\rho_{v}}{m_{v}}+\frac{\rho_{b}}{m_{b}}\right)kT$. 
In energy equation the transfer of energy between vapor and background gas species
are considered due to the heat flux $\vec{q}$, stress tensor $\bar{\bar{Q}}$ and 
also due to the different value of energy contents in the vapor and background 
gas species. The quantity $\vec{q}$ in Eq. (4) is the heat flux defined as,
$\vec{q}=-k_{mix}\vec{\nabla} T$ where $k_{mix}$ is the thermal conductivity of
the binary gas mixture. The thermal conductivity of individual species is
determined using the semi empirical formulation and this is given by the 
following expression:
\begin{eqnarray}
k_{i}=\frac{1}{d_{i}^{2}}\sqrt{\frac{k^{3}T}{\pi^{3}m_{i}}}
\end{eqnarray}
\subsection{Simulation, melting and vaporization of the target material:}
Experimental study suggest that the time evolution of plasma plume in the background gas is the axis 
symmetry phenomena. So to avoid the mathematical complexity and also the computational expenses required 
in order to perform the numerical study in three-dimension (3D), we carry out simulation in two-dimension 
(2D). Therefore we solved Eqs. $\left(1-4\right)$ numerically in two-dimension (2D) using the flux 
corrected scheme of Boris {\it et. al.} \cite{boris1993}. \par
In figure $1$ we displayed the schematic diagram of 2D computational domain and boundary conditions 
imposed on the boundaries. We used solid wall and inflow boundary conditions for the boundary. In solid 
wall boundary conditions the velocity component normal to boundary is taken to be zero and also the 
gradient of the quantities pressure $p$, transverse component of velocity $v_{y}$ and density $\rho_{v}$ 
normal to the boundary are set to zero. In inflow boundary condition the normal component of the flow 
velocity $v_{x}$ at the boundary should be constant, pressure and density is also constant at the 
boundary. Boundaries assigned by the numbers $1,3,4,5,6,7$ is kept at the solid wall boundary condition 
through the simulation. Region 2 are the locations where the plasma plume will be produced throughout the 
ablation process. So region two is kept at inflow boundary boundary condition during the ablation process. 
When the plumes are completely formed in the computational space, region 2 is switched with the solid wall 
boundary condition. Time required in the ablation process is computed using the formula, $t_{vap}=
\frac{M_{a}}{\rho_{v}v_{v}A}$ where $M_{a}$ is the mass of the ablated thin film which can be evaluated 
from the density of the film, its thickness and the cross-section area A of the focused laser spot. \par
Surface temperature $T_{s}$ of the ablated material is calculated by equating the 
incident laser energy with the energy required for the melting of the target material and further its 
vaporization. 
\begin{eqnarray}
F_{L}A(1-
R)=M_{A}c_{S}{(T_{boi}-T_{r})+(T_{S}-
T_{boi})}+M_{A}\lambda{e}
\end{eqnarray}  
or,
\begin{eqnarray}
T_{s}=T_{r}+\frac{F_{L}A\left(1-R\right)-M_{A}\lambda_{e}}{c_{s}M_{A}}
\end{eqnarray} 
Where $F_{L}$ is the laser fluence, $A$ is the spot size of the laser, 
$R$ is the reflectivity, $c_{S}$ is the specific heat of material, $
\lambda_{e}$ is the latent heat of evaporation, $T_{boi}$ is the 
boiling temperature and $T_{r}$ is the room temperature. Further we 
use $T_{s}$ for the computation of the surface pressure $p_{s}$ using 
the the Clausius-Clapeyron equation \cite{bogaerts2003},
\begin{eqnarray}
p_{s}=p_{0}exp\left[\frac{M_{A}\Delta H_{vap}}{k_{b}}\left\lbrace
\frac{1}{T_{boi}}-\frac{1}{T_{s}}\right\rbrace\right]
\end{eqnarray}
Here $T_{boi}$ is normal boiling temperature at standard pressure 
$p_{0}$ and $\Delta H_{vap}$ is the vaporization enthalpy at 
temperature of $T_{boi}$ at standard pressure $p_{0}$.
The ablated material from the surface moves towards the ambient gas and the
Knudsen layer is formed whose thickness is of the order of a few mean free paths. 
In the Knudsen layer there is a drop in both temperature as well as the pressure
of the ablated material by the following factors:
\begin{eqnarray}
T=0.67T_{s}; p=0.21p_{s}
\end{eqnarray}
Our computational domain starts after this Knudsen layer. Therefore we use Eq. $\left(13\right)$
for the estimation of the temperature $T$ and pressure $p$ of the inflowing
ablated material. These then decide density $\rho_{v}$ of the incoming ablated
material (used in Eq. $\left(14\right)$) from the ideal gas law. The velocity with which 
the vapor enters the ambient medium is approximated by the velocity of sound.
\begin{eqnarray}
v_{v}=\left(\frac{\gamma k_{b}T}{m_{v}}\right)^{1/2}
\end{eqnarray}
\begin{figure}[h]
\begin{center}
\includegraphics[width=10cm]{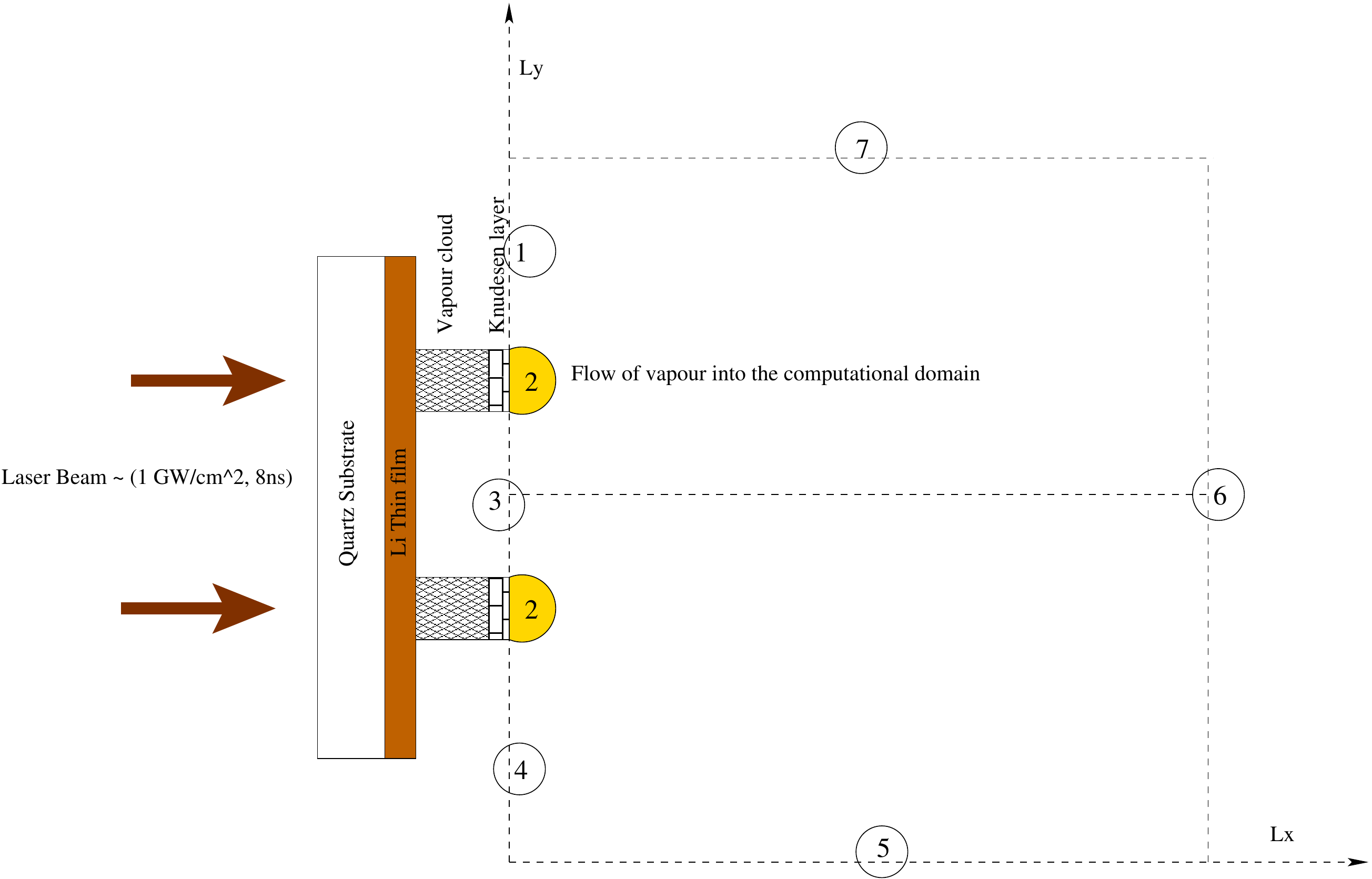}
\caption{Schematic diagram of 2D computational space in the ${\it xy}$ plane; each of sides $L_{x}$ and $L_{y}$ are $2.5 m$ span in the region from $0$ to $2.5$;
we also display the relevant components used in the experiment though not directly 
involved in the simulation. Numbers inside the circle represent the different boundary 
conditions imposed at boundaries. These boundary conditions are as follows: 
$\textcircled{1}   
  v_{x}=0,\frac{\partial p}{\partial x}=\frac{\partial \rho}{\partial x}=\frac{\partial 
v_{y}}{\partial x}=0$; 
 $\textcircled{2} \left( t > t_{vap} 
\right) v_{y}=0,p=p,\rho=\rho_{v},v_{x}=v_{v}$; 
 $\textcircled{2} \left( t < t_{vap} \right) v_{x}=0,\frac{\partial p}{\partial x}=\frac{\partial \rho}{\partial x}=\frac{\partial v_{y}}{\partial x}=0$;
 $\textcircled{3} v_{x}=0;\frac{\partial p}{\partial 
x}=\frac{\partial \rho}{\partial x}=\frac{\partial v_{y}}{\partial x}=0$; 
 $\textcircled{4} v_{x}=0;\frac{\partial p}{\partial x}=\frac{\partial \rho}{\partial x}=\frac{\partial v_{y}}{\partial x}=0$;
 $\textcircled{5} v_{y}=0;\frac{\partial p}{\partial y}=\frac{\partial 
\rho}{\partial y}=\frac{\partial v_{x}}{\partial y}=0$; 
 $\textcircled{6} v_{x}=0;\frac{\partial 
p}{\partial x}=\frac{\partial \rho}{\partial x}=\frac{\partial v_{y}}{\partial x}=0$;
 $\textcircled{7} v_{y}=0;\frac{\partial p}{\partial y}=\frac{\partial \rho}{\partial 
y}=\frac{\partial v_{x}}{\partial y}=0$; Note that this figure is borrowed from our earlier work \cite{yadav2017} 
}
\end{center}
\end{figure}
%
The values of the physical properties of the carbon target material used in the simuation are as follows: $\rho_{s}=1.4$ $g/cm^{3}$; $T_{boi}=4492^{o}C$; $c_{s}=710 $ $J/kg.K$; $\lambda_{e}=356 $ $kJ/mol$; $R=0.5$, where $\rho_{s}$, $T_{boi}$, $c_{s}$, $\lambda_{e}$ and $R$ are respectively, bulk density, boiling point, specific heat, latent evaporation and reflectivity. 
The initial values
of various parameters/quantitites (required in the simulation) are computed using formula given in this section. The value of these physical parameters/quantities are as follows: temperature
$T=9.35\times 10^{4}$ $K$, pressure $p=1.25\times 10^{7}$ $Pa$, density $\rho=1.93$ $kg/m^{3}$
, velocity $v_{v}=9.53\times 10^{3}$ $m/s$, vapor-flux $J=1.84 \times 10^{4}$ $kg/m^{2}.s$ and vaporization-time $t_{vap} = 3.8\times 10^{8}$ $
s$.
\section{ Numerical Results and Discussion}
We solved numerically Eqs. $\left(1-4\right)$ for the study of the lateral interaction of two plasmaplumes in the 
presence of the various background gases. The background gases considered in the simulations are $He$,
$Ne$, $Ar$ and $Xe$. The separation between the plumesare considered $\sim 5 mm$. The simulations are 
performed at two background pressures $1$ and $3$ \hspace{0.1cm} $mbar$ in the all cases. \par
Time evolution of the plumes at $1$ $mbar$ and $3$ $mbar$ background pressure in the presenceof $He$, $Ne
$, $Ar$ and $Xe$ are shown in Figures $2$ and $3$ respectively. Visible examination of the images in 
figure $2$ clearly show the effect of mass of ambient gas on the dynamics of expanding plasma plumes. In 
case of lighter gas i.e. $He$, plasma plume expands linearly up to the considered time delay as similar to 
the expansion in vacuum. An additional fable semicircular structure appeared ahead of the plasma plume at 
time delay $508$ $ns$ which is attributed as shock front. Here, the sound wave velocity 
$c_{s}$ in $He$, $Ne$, $Ar$ and $Xe$ are $10.22\times 10^{2}$ $m/s$, $4.54\times 10^{2}$ $m/s$, $3.23\times 
10^{2}$ $m/s$ and $1.78\times10^{2}$ $m/s$ respectively. Using relation $\left(14\right)$ the estimated 
initial plume velocity is $\sim 1.12\times 10^{4}$ $m/s$, which is much higher than the sound wave velocity 
in the considered medium. This satisfy the strong shock condition especially in heavier background gas. 
The shock fronts are different to each other in terms of size and intensity depending upon background gas 
used in the simulation. In case of $Ne$, shock front is clearly visible at $304$ $ns$ whereas it appears 
further earlier $\sim 208$ $ns$ in the case of heavier ambient gases that is $Ar$ and $Xe$. Also the shock 
front is strengthened and more intense in heavier background gas. \par
In order to further study of the dynamics of plasma induced shock waves, axial-plume-dimensions (distance 
from the target to shock front) are analysed as a function of time. Figure $5$ shows the variation of the 
distance of the shock-front from the target (also known as shock-distance) with time for all the cases of 
the simulation of the ambient gas performed at $1$ $ mbar$ pressure. \par
In all cases we observe that initially for a very short time period, plasma plume expands linearly in all 
background gases. After a certain time delayed depending upon the mass of the background gas, the 
expansion of the plume is deviated from the linear expansion and it follows $ \sim t^{0.4}$ dependence 
(see figure $5$). It is clearly visible in our simulation as shown in figure $2$, $3$ and  
$4$ where in heavier background gas shock wave initiated earlier in comparison to that observed in the 
case of the simulation of lighter-gas at a fixed background pressure. Delayed non-linear expansion 
behavior and its dependence on the mass of the background gas is in agreement with the 
blast wave model that describe the expansion of the massive-shock present in the gases system. 
In figure $6$ we observe that shock front appears roughly at 
$0.012$ $m$, $0.009$ $m$, $0.008$ $m$ and $0.006$ $m$ from the target surface for the case $He$, $Ne$, $Ar$ and $Xe$ respectively at time $t=508.0$ $ns$. The limiting characteristic distance $\left(R\right)$ for the shock front can be estimated 
by equating the mass of the gas encompassed by shock wave and initial abalted mass $\left(M_{a}\right)$, that 
is $R = (3M_{a}/2\pi\rho_{b})^{1/3}$ where $\rho_{b}$ is the density of background gas. The estimated  distance of the shock front
$\left(R\right)$ at which the shock wave model is valid for $1.0$ $mbar$ pressure of $He$, $Ne$, $Ar$ and $Xe$ are $0.018$ $m$, $0.01$ $m$, $0.0083$ $m$ and $0.0057$ $m$
respectively. Hence our simulation roughly capture the limiting characteristics 
distance (plume length) for shock wave expansion as shown in Figure $6$. \par
In order to get more insight into the intensity jump at the shock 
front and its strength in different background gases, the intensity profiles along 
the expansion axis are analyzed at onset of time.  \par 
Variation of intensity/density in the plasma plume at $1$ $mbar$
pressure and $t = 508.0$ $ns$ along the expansion axis ($x-$axis) and at $y \sim 0.15$ $m$
for all the cases of simulation of background gases is shown in Figure $6$. 
The density jump ahead of the plasma plume due to shock wave 
formation and its delayed appearance with the decrease of the mass of background gas 
is clearly visible in Figure $6$. Also the intense and sharp intensity jump in the 
simulation of $Xe$ background gas in comparison to the $He$ background gas reveals
that shock strength (reciprocal of shock thickness) is higher in case of 
heavier background atoms/gas. \par
Further, simulated results depicted in Figures $2$ and $3$ are also compared with 
theoretically predicted shock-front velocity and density. 
According to the theory, the shock wave parameters especially the shock-front density and
shock-front velocity could be determined based on the mass and energy conservation formula
as given below:
\begin{eqnarray}
V_{s}\approx \left(\frac{\gamma +1}{2}\right)V_{a}
\end{eqnarray}
\begin{eqnarray}
\rho_{s}\approx \left(\frac{\gamma+1}{\gamma-1}\right)\rho_{b}
\end{eqnarray}
Where $V_{s}$ and $V_{a}$ are the shock-front velocity and velocity of the plume during the initial time 
evolution of the plume respectively. $\gamma$ is the ratio of specific heats at constant pressure and 
volume. As, in our simulation we are using the monatomic gas so the value of $\gamma$ is $1.67$. $\rho_{s}
$ and $\rho_{b}$ are the shock front density and background density respectively. Thus according to the a
above formula if massive shock is present in the system then the shock density should be roughly $\sim 
4.0\rho_{b}$ and shock wave velocity, $V_{s}\sim 1.34 V_{a}$. The simulated density at the 
shock fronts in the case of $He$, $Ne$, $Ar$ and $Xe$ are $2.5\times 10^{-3}, 8.3\times 10^{-3}, 
1.25\times10^{-2} $ and $2.0\times10^{-2}$ respectively in the dimensionless unit (see, Figure $6$). In 
the simulation, the density is normalized by the plume-density $\rho_{v}$ at $t=0.0$ that is $\rho_{v}
^{t=0.0}=1.75$ $kg/m^{3}$. Thus the actual value of density in the shock fronts (at $t=304.0 ns$ and at $1$ 
$mbar$ pressure) are as follows, $\rho_{v}^{He}= 4.38 \times 10^{-3}, \rho_{v}^{Ne}=1.46\times10^{-2}, 
\rho_{v}^{Ar}=2.2\times 10^{-2}$ and $\rho_{v}^{Xe}=3.5\times 10^{-2}$ in the unit of $kg/m^{3}$. If one 
compare these values with its corresponding background gas density that is $\rho_{b}^{He} \sim 1.6\times 
10^{-4}$ $kg/m^{3}, \rho_{b}^{Ne}\sim 8.1\times 10^{-4}, \rho_{b}^{Ar}\sim 1.6 \times 10^{-3}$ and $\rho_{b}
^{Xe}\sim 5.3\times 10^{-3}$ (in the unit of $kg/m^{3}$), the value of shock front density is always 
greater than $\sim 10$ times of the background density in all cases of the simulation which is 
overestimated the theoretical value of the ideal blast wave model. \par
Further, we estimated the velocity $\left(V_{a}\right)$ of the ablated plume using the 
relation, $v_{v}=\left(\frac{\gamma kT}{m_{v}}\right)^{1/2}$. The obtained value of $V_{a}$ is $1.1\times 
10^{4} m/s$. This is the velocity with which plume enter the simulation space in the presence of 
background gas. Note that plume velocity $V_{a}$ does not change significantly during the initial time 
evolution of the plume in the background gas. Later when strong shock present in the system then the 
average shock velocity $V_{s}$ is determined using the fitting parameters $\alpha$ and $\beta$. $\alpha$ 
and $\beta$ are constants, appear in the function $\alpha t^{0.4}+\beta$ (also known as a Taylor Sedov 
model) that is used to fit the shock front position vs. time curve (see Figure $5$). The value of $\alpha$ 
and $\beta$ depends upon the mass of the background gas. In the case of the simulation of $Xe$ background 
gas, the estimated shock velocity $V_{s}$ is $\sim 1.69\times 10^{4}$ $m/s$. This value is approximately 
$1.5$ times of the initial plume velocity $V_{a}$ that is in good agreement with the 
theoretical approximation $\left(V_{s}/V_{a}=1.3\right)$. Values of $\alpha$ and $\beta$, and shock 
velocity $V_{s}$ are summarized in table $I$ for the simulation of all the ambient gases performed at $1$ 
$mbar$ pressure. \par
\begin{table}
\caption{Values of $\alpha$ and $\beta$, and shock velocity $V_{s}$ $\left(\sim 0.4\alpha t^{-0.6}\right)$ in simulations of $He$, $Ar$ and $Xe$ ambient gases at $1$ $mbar$ pressure.}
\begin{center}
\begin{tabular} {c c c c c}
\hline
\hline
{\bf Ambient gas} &
{\bf $\alpha$} &
{\bf $\beta$}  &
{\bf $V_{s}$ $(m/s)$} \\
\hline
 $He$  &
 $2.6557\times 10^{-3}$  &
 $-1.6067\times 10^{-2}$  &
 $2.55 \times 10^{4}$ \\
 $Ar$  &
 $2.1597 \times 10^{-3}$  &
 $-1.2596 \times 10^{-2}$  &
 $2.08 \times 10^{4}$ \\
 $Xe$  &
 $1.7586 \times 10^{-3}$  &
 $-9.977 \times 10^{-3}$  &
 $1.69 \times 10^{4}$ \\
\hline
\hline
\end{tabular}
\end{center}
\end{table}
Another noteworthy observation of the present simulation is that it predict the density 
discontinuity in between the plasma plume and shock front due to the hydrodynamic movement of plume and 
shock front as reported by several experimental works in the past. Two region of discontinuity is 
clearly visible in case of the simulation of the background gases $Ne$ and $Ar$ as shown in Figure 6. While 
in the case of the simulation of $Xe$ background gas this is not visible due to the nominal separation 
between the plume and shock-front. Also in the case of the simulation of $He$ background gas this 
discontinuity is not clearly visible, here it is because of the weak-shock condition (see Figure $6$). In 
most of the previous work, the first discontinuity is attributed to the ionized shocked gases just behind 
the shock front. Whereas the second discontinuity, known as contact surface (CS) which is the boundary of 
ablated species. Using the Westwood model \cite{westwood1978} we can easily understand the change in the direction of carbon species in a single collision with $Ne$ and $Ar$ background gas species.
Thus, elastic scattering between the leading plume species and heavier background atoms play 
the significant role in the build-up of the density in the CS region. \par
Apart from the characteristic expansion of the plasma plume and shock wave formation, our simulation also 
capture the most of the features of lateral interactions between two 
spatially separated plumes and its dependence on the background conditions (that is 
pressure and mass of the background gas.The earlier studies report the formation of well defined 
interaction region (or stagnation region) in parallel to propagating plasma plumes in close proximity.
Due to the angular distribution of the plume species, interaction between the counter 
propagating plume species with the condition $\eta = \frac{D}{\lambda} >> 1$, is responsible for the formation 
of interaction region. Where $\eta$, $D$ and $\lambda$ are respectively, the collisionality parameter, 
separation between the plumes and collision frequency. For the higher value of $\eta$, the counter 
propagation energetic particles lose energy rapidly due to multiple collisions and formed the interaction 
region. Thus the value of $\eta$ define whether the interaction between the plumes result into the 
interpenetration or it simply turned into the interaction zone (as observed in the simulation). \par
The one to one comparison between the images observed in different background gas and 
pressure (as shown in Figures $2$ and $3$) indicates that dynamics and geometrical structure of the 
expanding vapour, shock front and interaction zone at $1$ $mbar$ $Ne$ pressure is nearly similar to the 
simulated images at $3$ $mbar$ $He$ pressure. On the same way, the characteristics of the plumes and its 
interaction pattern at $1$ $mbar$ of $Xe$ pressure resembles that observed at $3$ $mbar$ $Ar$ pressure. 
Based on the above observations, the interaction mechanism between the plumes is broadly divided into two 
region that is in the region where absence or weak shock condition and presence of strong shock front.
In absence of shock front, for example at $208$ $ns$ and $304$ $ns$ in $He$ and at $208$ 
$n$s in $Ne$ background at $1$ $mbar$ pressure (see, figure $2$), an additional luminous components in 
between the plumes in our simulation is treated as conventional interaction region formed by the multiple 
collisions between the counter propagating species at the middle of two expanding plumes. \par
The present simulation is also predict the higher expansion velocity of interaction region in comparison 
to interacting plumes, which is in excellent agreement with previously reported experimental results. 
Scenario 
is different in presence of shock wave ahead of plume where shock-shock interaction and subsequent 
reflections come in picture along with the interaction between plume species. With the assumption of 
collision between two planner shock fronts, shock-shock interactions and its subsequent reflections is 
classified as regular and Mach reflections depending upon the shock strength and angle of incidence with 
respect to the plane of symmetry. \par
The carefull examination of the simulated images in case of $Ne$ and $Ar$  at $1$ $mbar$ pressure and at time delay $>200$ 
$ns$ (Figure $2$) and also at $3$ $mbar$ of $He$ pressure (Figure $3$), It can be clearly seen that shock fronts are 
physically interact at
the middle of the two plumes. In this simulation both plume is identical and therefore point of 
interaction is always lying in the middle and moving along the expansion axis with time as observed in 
Figure $2$. This good agreement with theoretically predicted interaction and regular reflection of two relatively 
weak shock fronts. Also the predicted structure of the interaction zone is an excellent agreement 
with the experimental observation in the similar condition. \par
On the other hand in case of heavier background gas ($Ar$ and $Xe$) and especially at higher pressure and 
later time delay where the strong shock is predicted, luminous interaction region is not observed in 
simulation even at highest considered time delay ( see, Figure $3$). Also in contrast to the weak shock 
condition, physical overlapping between the shock fronts is not observed in this region.  The density 
discontinuity in between the plumes is clearly visible in $3$ $mbar$ pressure of $Ar$ and $Xe$ at $500$ $ns$ time 
delay (Figure $3$) where the interacting shock wings is seems to repelled each other  and hence completely 
suppressed in opposite direction. This observation is resembled with the case of Mach reflection between 
the two interacting strong shock fronts. In this case point of reflection is split into two 
symmetrical points and move in opposite direction in plane perpendicular to expansion axis. The 
reflections from these two points restrict the escape of plume species and form the density discontinuity 
in between the plume as shown in Figure $3$. \par
Interaction between the plumes as well as shock fronts and its dependence on the shock strength is represented 
in better way by comparing the interactions of plasma plumes in $He$ and $Xe$ background gases at $3$ $mbar$ pressure as shown in 
Figure $4$. As discussed earlier, conventional interaction region is observed at $\sim 200$ $ns$ in $He$ 
where the shock front is insignificant. Even at $3$ $mbar$ $He$ pressure, relatively weak shock is 
predicted in our simulation. Thus the simulated structure of interaction region at $t > 200 $ $ns$ is in 
line with the interaction between two weak shock fronts. However the complete suppression of overlapping 
of plume species and shock wings and hence the formation of interaction zone because of Mach reflection 
between the strengthen shock front in $Xe$ background is correctly predicted in our simulation. \par
 In this simulation we successfully predicted the most of the 
features of colliding plumes and shock fronts and also the transition from regular to Mach shock 
reflection in reference of weak and strong shock conditions.
\begin{figure}[h]
\begin{center}
\includegraphics[width=14cm]{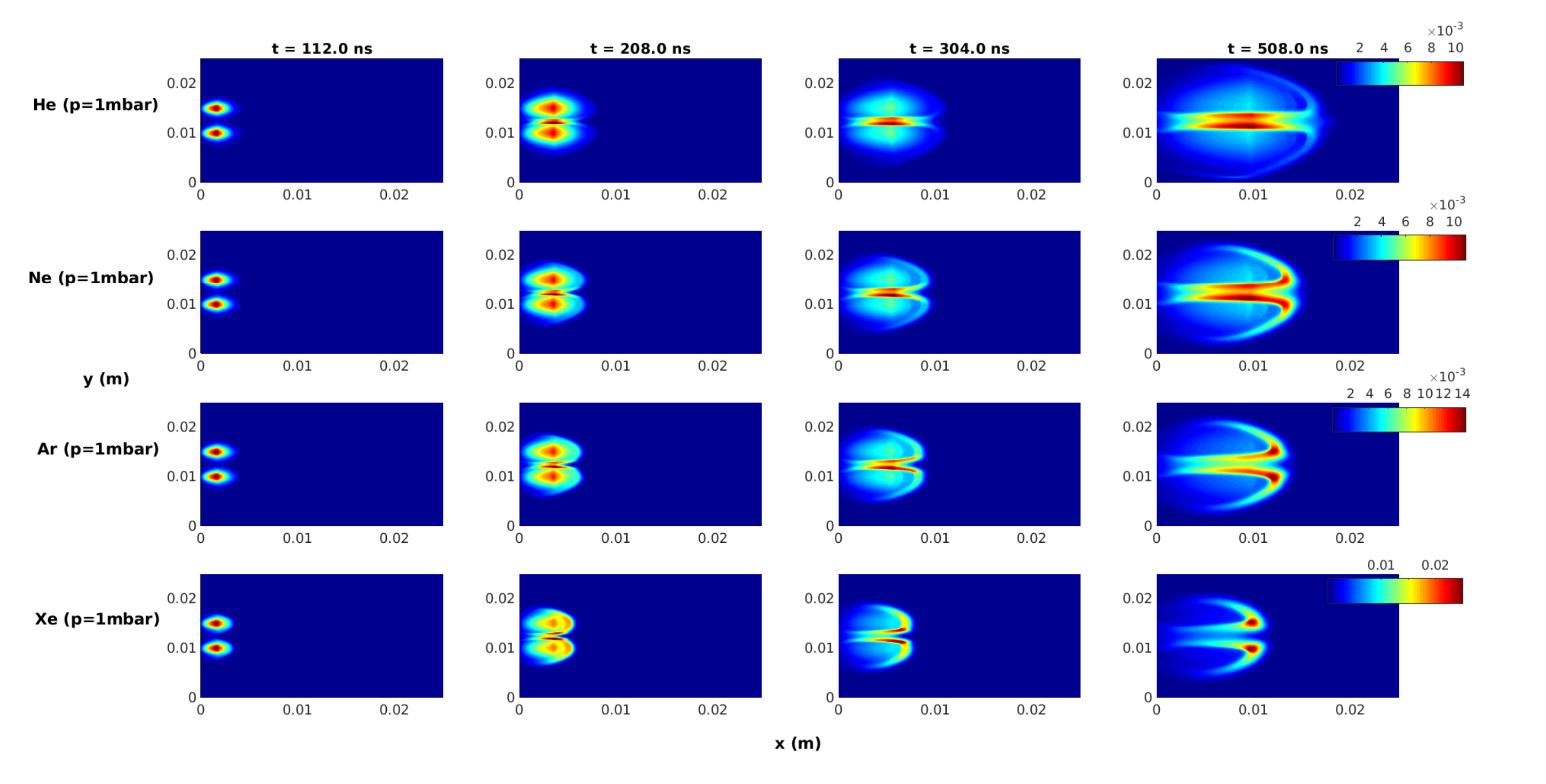}
\caption{
(left to right) Time evolution of two plasma plumes from simulations
in the presence of $He$, $Ne$, $Ar$ and $Xe$
background gases at $1.0$ $mbar$ pressure; initial separation between plumes is considered
$5.0$ $mm$.}
\end{center}
\end{figure}
\begin{figure}[h]
\begin{center}
\includegraphics[width=14cm]{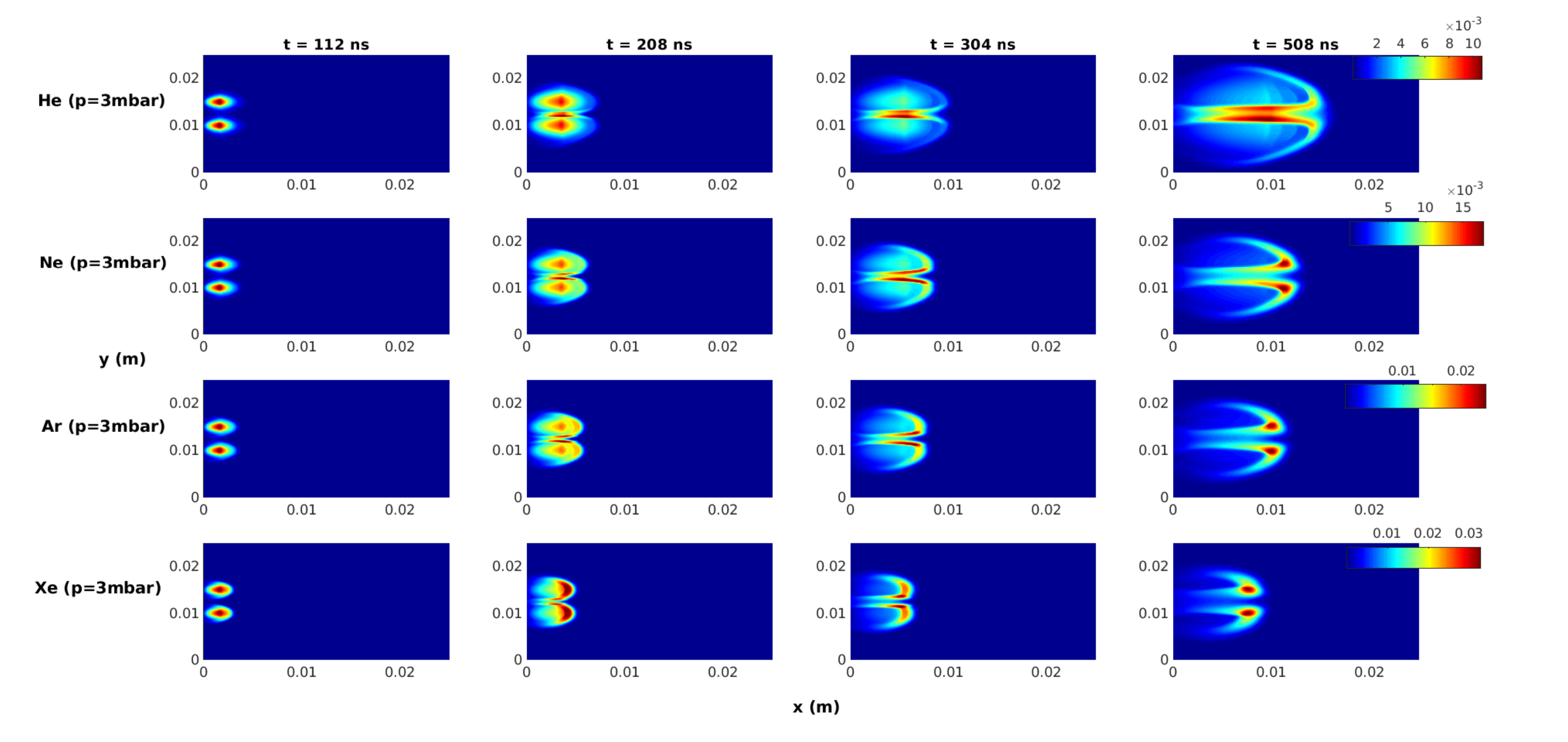}
\caption{
(left to right) Time evolution of two plasma plumes from simulations in the presence of $He$, $Ne$, $Ar$ and $Xe$
background gases at $3.0$ $mbar$ pressure; initial separation between plumes is considered
$5.0$ $mm$.}
\end{center}
\end{figure}
\begin{figure}[h]
\begin{center}
\includegraphics[width=12cm]{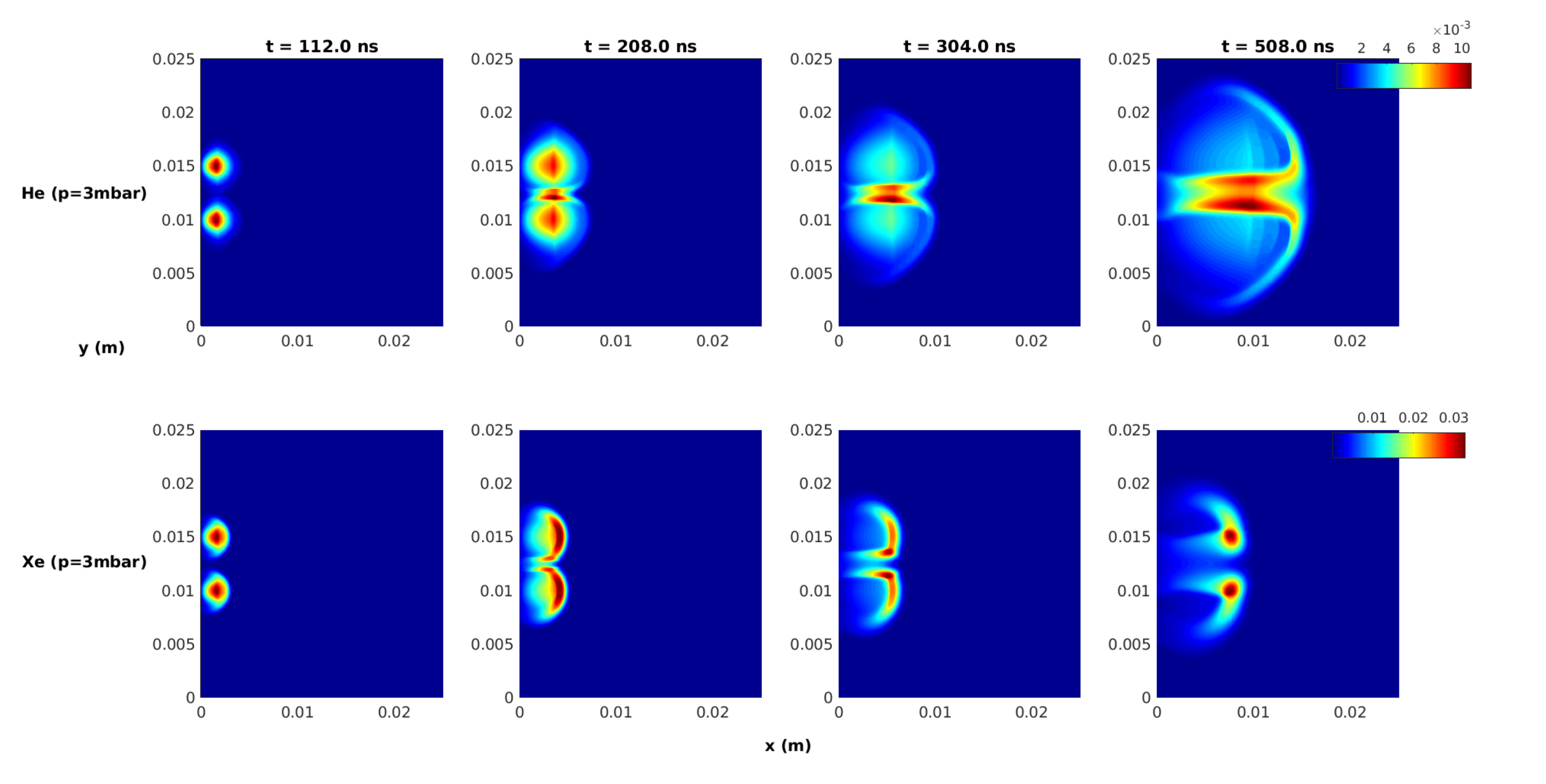}
\caption{Comparison of the images of the time evolution of plasma plumes from simulations in $He$ and $Xe$ background gases at $3.0$ $mbar$ pressure.}
\end{center}
\end{figure}
\begin{figure}[h]
\begin{center}
\includegraphics[width=12cm]{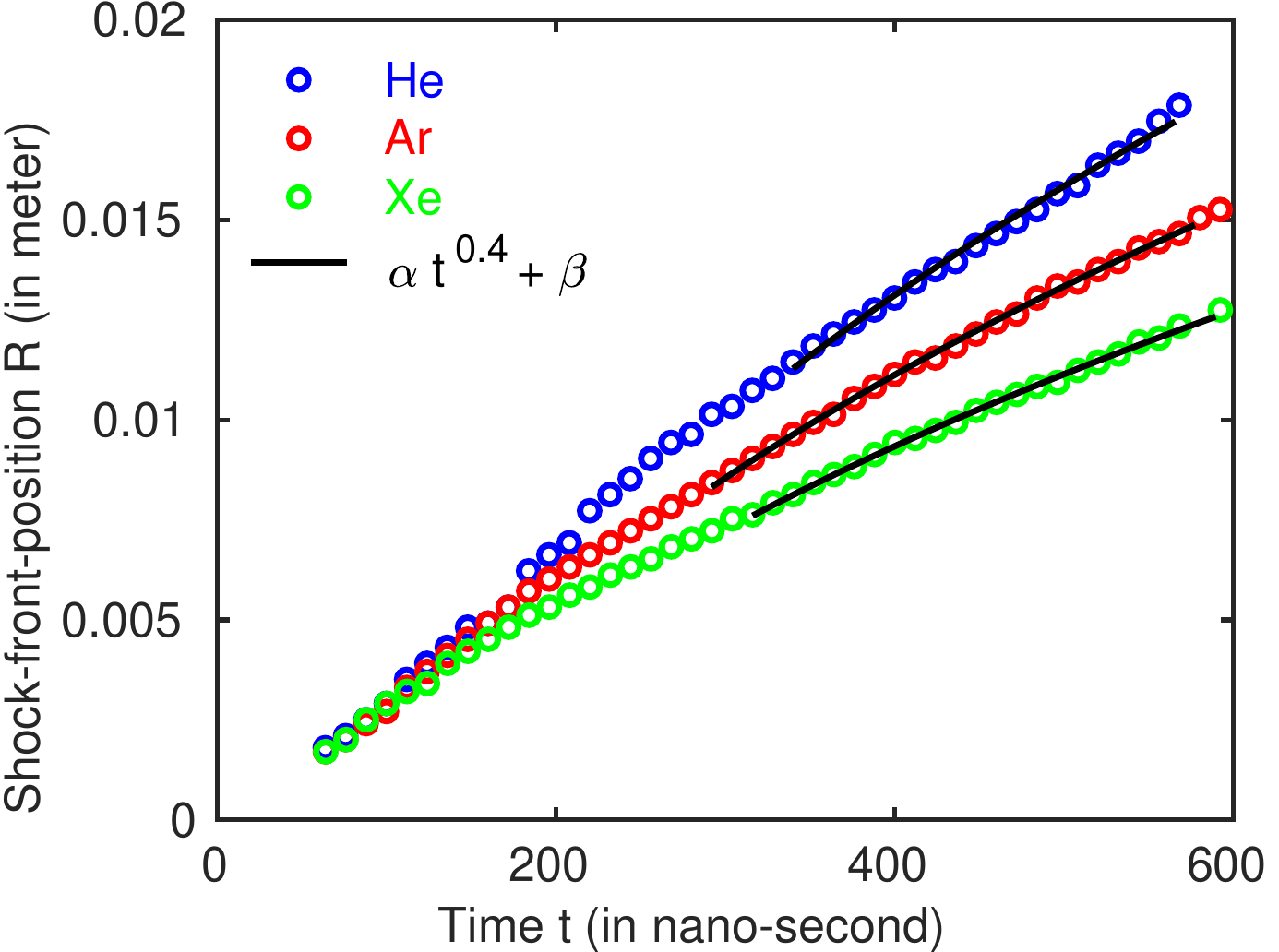}
\caption{
Plot of the shock front position versus time in the simulations of $He$, $Ar$
and $Xe$ background gases at $1.0$ $mbar$ background pressure. Each of them is fitted
at later time beyond $\sim 300$ $ns$ using function, $\alpha t^{0.4} + \beta$ where $\alpha$ and $\beta$ are the fitting
constants, and its value depends upon the background gas and pressure. The value of
$\alpha$ and $\beta$ are summarized in table $I$.}
\end{center}
\end{figure}
\begin{figure}[h]
\begin{center}
\includegraphics[width=12cm]{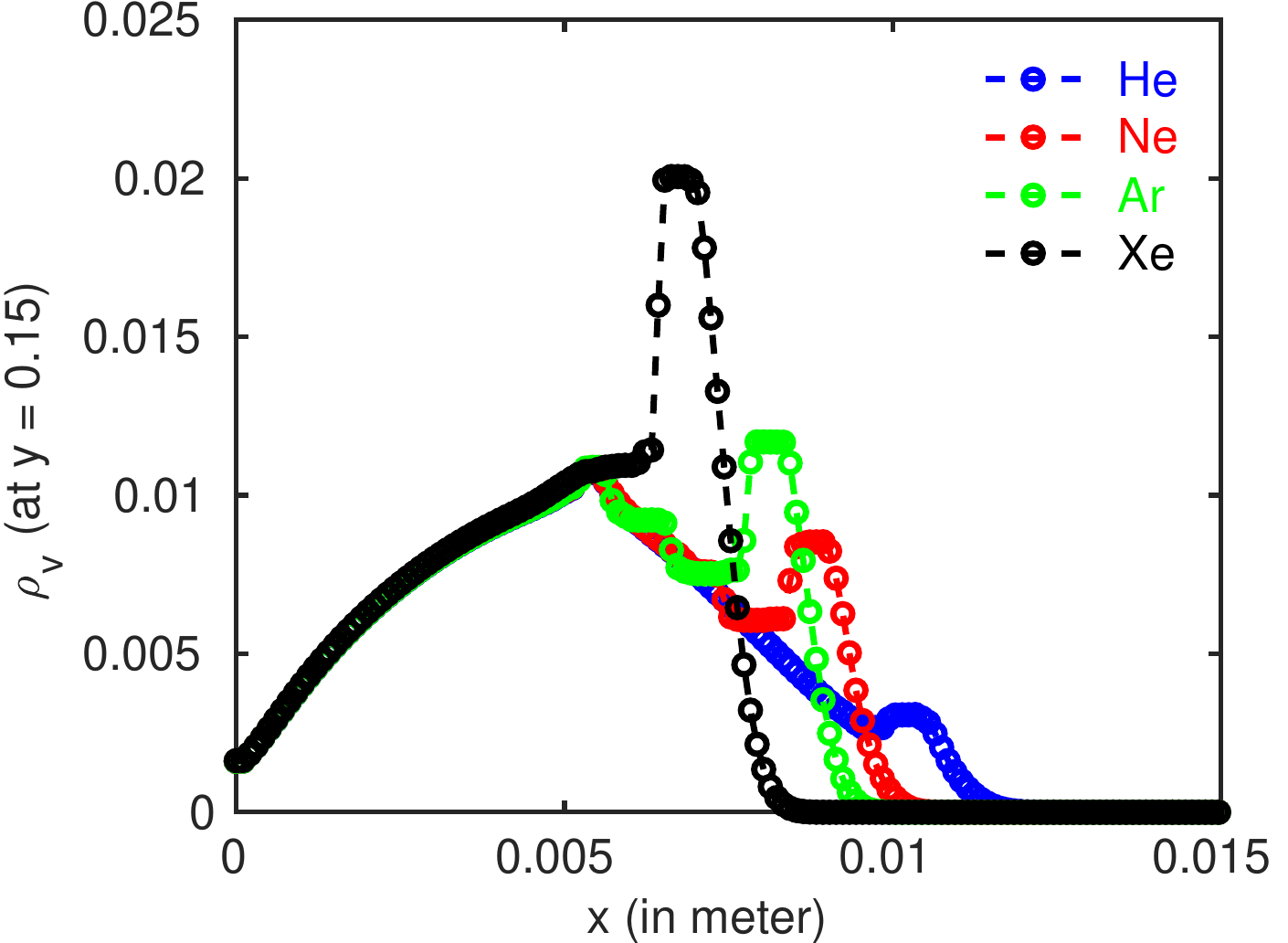}
\caption{
Variation of density (and/or intensity) along $x-$ axis in the upper plume of
the image displayed in figure $2$ at $t = 500.0$ $ns$ from the simulations of background
gases, $He$, $Ne$, $Ar$ and $Xe$ at $1.0$ $mbar$ background pressure. The variation of density
is passing through the center of the plume (located at $y = 0.15$ $m$ position).}
\end{center}
\end{figure}
\section{Conclusion}
In this work we presented the numerical simulation 
of the evolution of two spatially separated plasma plumes in the presence of $He$, $Ne$, 
$Ar$ and $Xe$ background gases at $1$ and $3$ $mbar$ pressures. The features of expanding 
plasma plumes in close proximity and interaction between them in different background gas 
captured in the present simulation are in close agreement with the reported experimental 
observations \cite{kumar2016}. Effect of mass of ambient gas on plume dynamics, initiation of shock waves, 
shape and strength of shock front and variation of the density/intensity in the plume as well as shock region 
is precisely reproduced in the simulation. Further lateral interaction between the two 
expanding plumes and formation of interaction region are also predicted well by the 
simulations. The structure formation because of the shock-shock interaction and reflection 
observed in simulation is in good agreement with the experimental results. The observed 
results suggest the presence of the regular shock-reflection in the cases of the simulation of $He$ and $Ne$ background gases 
and Mach-reflection in the cases of $Ar$ and $Xe$ simulations especially 
at later time. The simulation in the different ambient gases demonstrate the 
transition from regular to Mach reflections of shock wave depending upon the shock 
strength.

\section*{References}
\begin{thebibliography}{10}
\bibitem{allen1995}
M. von Allen and A. Blatter, Laser-Beam Interactions with Materials:Physical Principles and Applications (Springer, Berlin, 1995).
\bibitem{chrisey1994}
D. B. Chrisey and G. K. Hubler, Pulsed Laser Deposition of Thin Films (Wiley, New York, 1994).
\bibitem{choudhury2019}
Kaushik Choudhury, R.K. Singh, P. Kumar, Mukesh Ranjan, Atul Srivastava and Ajai Kumar, Nano-Structures $\&$ Nano-Objects {\bf 17}, 129  (2019).
\bibitem{miziolek2006}
A. W. Miziolek, V. Palleschi, I. Schechter, Laser Induced Breakdown Spectroscopy (Cambridge University Press, 2006).
\bibitem{freeman2011}
J. R. Freeman, S. S. Harilala), and A. Hassanein, Journal of Applied Physics {\bf 110}, 083303 (2011).
\bibitem{zakharov2003}
Y. P. Zakharov, IEEE Transactions on Plasma Science {\bf 31}, 1243 (2003).
\bibitem{huber2005}
A. Huber, U. Samm, B. Schweer, and Ph. Mertens, Plasma Phys. Controlled Fusion {\bf 47}, 409 (2005).
\bibitem{hough2010}
P. Hough, C. McLoughlin, S. S. Harilal, J. P. Mosnier and J. T. Costello, J. Appl. Phys. {\bf 107}, 024904 (2010).
\bibitem{bulanov2002}
S. V. Bulanov, T.Z. Esirkepov, F.F. Kamenets, Y. Kato, A.V. Kuznetsov, K. Nishihara, F. Pegoraro, F. Tajima, and V. S. Khoroshokov, Plasma Phys. Rep. {\bf 28}, 975 (2002).
\bibitem{rancu1995}
O. Rancu, P. Renaudin, C. Chenais-Popovics, H. Kawagashi, J.C. Gauthier, M. Dirksmeoller, T. Missalla, I. Uschmann, E. Forster, O. Larroche, O. Peyrusse, O. Renner, E. Krousky, H. Pepin, and T. Shepard, Phys. Rev. Lett. {\bf 75}, 3854 (1995).
\bibitem{wan1997}
A. S. Wan, T. W. Barbee, R. Cauble, P. Celliers, L. B. Da Silva, J. C. Moreno, P. W. Rambo, G. F. Stone, J. E. Trebes, and F. Weber, Phys. Rev. E {\bf 55}, 6293 (1997).
\bibitem{gregory2008}
C. D. Gregory, J. Howe, B. Loupias, S. Myers, M. M. Notley, Y. Sakawa, A. Oya, R. Kodama, M. Koenig, and N. C. Woolsey, Astrophys. J. {\bf 676}, 420 (2008).
\bibitem{kuramitsu2011}
Y. Kuramitsu, Y. Sakawa, T. Morita, C. D. Gregory, J. N. Waugh, S. Dono, H. Aoki, H. Tanji, M. Koenig, N. Woolsey, and H. Takabe, Phys. Rev. Lett. {\bf 106}, 175002 (2011).
\bibitem{elton1994}
R. C. Elton, D. M. Billings, C. K. Manka, H. R. Griem, J. Grun, B. H. Ripin, and J. Resnick,  Phys. Rev. E {\bf 49}, 1512–1519 (1994).
\bibitem{dardis2010}
J. Dardis, J.T. Costello, Spectrochimica Acta Part B {\bf 65}, 627 (2010).
\bibitem{alshboul2014}
K. F. Al-Shboul, S. S. Harilal, S. M. Hassan, A. Hassanein, J. T. Costello, T. Yabuuchi, K. A. Tanaka, and Y. Hirooka, Phys. Plasmas {\bf 21}, 013502 (2014).
\bibitem{ake2006}
C. S$\acute{a}$nchez Ak$\acute{e}$, R. Sangin$\acute{e}$s de Castro, H. Sobral and M. Villagr$\acute{a}$n-Muniz. J. Appl. Phys. {\bf 100}, 053305 (2006).
\bibitem{harilal2011}
S. S. Harilal, M. P. Polek, and A. Hassanein, IEEE Trans. Plasma Sci. {\bf 39}, 2780 (2011).
\bibitem{eagleton1997}
R. T. Eagleton, J. M. Foster, P. A. Rosen, and P. Graham, Rev. Sci. Instrum. {\bf 68}, 834 (1997).
\bibitem{luna2007}
H. Luna, K. D. Kavanagh, and J. T. Costello, J. Appl. Phys. {\bf 101}, 033302 (2007).
\bibitem{kumar2014}
Bhupesh Kumar, R. K. Singh, Sudip Sengupta, P. K. Kaw, and Ajai Kumar, Phys. Plasmas {\bf 21}, 083510 (2014).
\bibitem{camps2002}
E. Camps, L. Escobar-Alarcón, E. Haro-Poniatowski, M. Fernández-Guasti, Appl. Surface Sci. {\bf 9}, 239 (2002).
\bibitem{hirooka2011}
Y. Hirooka, T. Oishi, H. Sato, and K. A. Tanaka, Fusion Sci. Technol. {\bf 60}, 804 (2011).
\bibitem{kumar2015}
 Bhupesh Kumar, R. K. Singh, Sudip Sengupta, P. K. Kaw, and Ajai Kumar, Phys. Plasmas {\bf 22}, 063505 (2015).
\bibitem{kumar2016}
Bhupesh Kumar, R. K. Singh, Sudip Sengupta, P. K. Kaw, and Ajai Kumar, Phys. Plasmas {\bf 23}, 043517 (2016).
\bibitem{singh2007}
R. K. Singh, Ajai Kumar, B. G. Patel and K. P. Subramanian, J. Appl. Phys. {\bf 101}, 103301 (2007).
\bibitem{lie1984}
Y. T. Lie, A. Pospieszczyk, and J. A. Tagle, Fusion Technol. {\bf 6}, 447 (1984).
\bibitem{pospieszczyk1989}
A. Pospieszczyk, F. Aumayr, E. Hintz, and B. Schweer, J. Nucl. Mater. {\bf 574}, 162–164, (1989).
\bibitem{kumuduni1993}
W. K. Kumuduni, Y. Nakayama, Y. Nakata, T. Okada and M. Maeda, J. Appl. Phys. {\bf 74}, 7510 (1993).
\bibitem{george2013}
Sony George, R.K. Singh, V.P.N. Nampoori and Ajai Kumar, Physics Letters A {\bf 377}, 391 (2013).
\bibitem{zeldovich2002}
Y. B. Zeldovich and Y.P. Raizer, Physics of Shock Wavesand High Temperature Hydrodynamic Phenomena (NewYork: Dover, 2002).
\bibitem{yadav2017}
Sharad K Yadav, Bhavesh G Patel, R. K. Singh, Amita Das, Predhiman K Kaw and Ajai Kumar, J. Phys. D: Appl. Phys. {\bf 50}, 355201 (2017).
\bibitem{patel2012}
Bhavesh G. Patel, Amita Das, Predhiman Kaw, Rajesh Singh, and Ajai Kumar, Phys. Plasmas {\bf 19}, 073105 (2012).
\bibitem{bird2006}
R. B. Bird, W. E. Stewart, and E. N. Lightfoor, {\it Transport Phenomena} ( Wiley 
2006).
\bibitem{bogaerts2003}
A. Bogaerts, Z. Chen, R. Gijbels, and A. Vertes,”Laser ablation for analytical sampling: what we
can learn from the modelling ?” Spectrochem. Acta. Part B {\bf 58 (11)}, 1867-1893 (2003).
\bibitem{boris1993}
J. P. Boris, A. M. Landsberg, E. S. Oran, and J. H. Gardner,"LCPFCT-Flux-corrected transport algorithm for solving generalized continuity equations" {\it Technical Report No. NRL/MR/6410-93-7192} Naval Research Laboratory, 1993.
\bibitem{westwood1978}
W. D. Westwood, Journal of Vacuum Science $\&$ Technology {\bf 15}, 1 (1978).
%
\end{thebibliography}
\end{document}